\documentclass[prl,showpacs,preprintnumbers,floatfix]{revtex4}
\usepackage{graphics}
\def \beq{\begin{equation}}
\def \eeq{\end{equation}}
\def \beqa{\begin{eqnarray}}
\def \eeqa{\end{eqnarray}}
\def \snn{\sqrt{S_{\scriptscriptstyle NN}}}
\def \ie{{\sl i.e.\/}}
\def \etal{{\sl et al.\/}}
\def \jhep{{\sl J.\ H.\ E.\ P.\ }}

\def \np{{\sl Nucl.\ Phys.\ }}
\def \pl{{\sl Phys.\ Lett.\ }}
\def \pr{{\sl Phys.\ Rev.\ }}
\def \prl{{\sl Phys.\ Rev.\ Lett.\ }}

\begin{document}
 
\title{Lattice QCD predictions for shapes of event distributions\\
  along the freezeout curve in heavy-ion collisions}
\author{R.\ V.\ \surname{Gavai}}
\email{gavai@tifr.res.in}
\affiliation{Department of Theoretical Physics, Tata Institute of Fundamental
         Research,\\ Homi Bhabha Road, Mumbai 400005, India.}
\author{Sourendu \surname{Gupta}}
\email{sgupta@tifr.res.in}
\affiliation{Department of Theoretical Physics, Tata Institute of Fundamental
         Research,\\ Homi Bhabha Road, Mumbai 400005, India.}

\begin{abstract}
We present lattice QCD results along the freezeout curve of heavy-ion
collisions. The variance, skew and kurtosis of the event distribution of
baryon number are studied through Pad\'e resummations. We predict smooth
behaviour of three ratios of these quantities at current RHIC and future
LHC energies. Deviations from this at lower energies signal the presence
of a nearby critical point.
\end{abstract}
\preprint{TIFR/TH/10-01}
\maketitle

All matter that we know undergoes phase transitions as external conditions
change. Strongly interacting matter, described by QCD, is not thought
to be exceptional.  Although the phase diagram of QCD matter is not
yet established, it has been the subject of intense theoretical and
experimental scrutiny in recent years.  Finding a critical point in this
system would be major landmark. One idea is to look at quantities which
are non-monotonic near a critical point. In this letter we give lattice
predictions of a few such observables;  we give precise results for the
backgrounds. These are the first lattice QCD predictions of
any quantity along the freezeout curve of heavy-ion collisions.
Any non-monotonic behaviour over these is a signal for the critical point;
and the lattice results near the critical point do show such behaviour.

Twenty years ago it became clear that QCD has no finite temperature
($T$) phase transition \cite{xov}, since all quarks are massive, as
evidenced by a non-vanishing pion mass. Lattice computations showed
that there remained a fairly abrupt change in thermodynamic properties
which could possibly influence the physics observed at the RHIC and
LHC. About a decade back it was realized that QCD could nevertheless
have a critical point at finite baryon chemical potential ($\mu_B$)
\cite{cp}. There is evidence for such a critical point from modern
lattice computations with two light dynamical flavours \cite{nt4,nt6},
and from an earlier small volume computation \cite{fk}.  The behaviour
of QCD with two light and a single heavier flavour is not yet settled,
although evidence for a critical point is mounting \cite{confused}.

One question of interest is whether this critical point can be found
in experiment. The results of \cite{nt4,nt6} suggest that heavy-ion
collision experiments at moderate colliding energies, achievable at 
RHIC, FAIR and NICA could find it. In these collisions of heavy-ions,
a fireball is created which evolves chemically and thermally before
freezing out. The primary question of experimental interest is how
close the freezeout is to the critical point. In this work we employ
Pad\'e resummation of lattice results to obtain predictions which
can be directly compared with experiments.

\begin{figure}[htb]
\begin{center}
   \scalebox{0.6}{\includegraphics{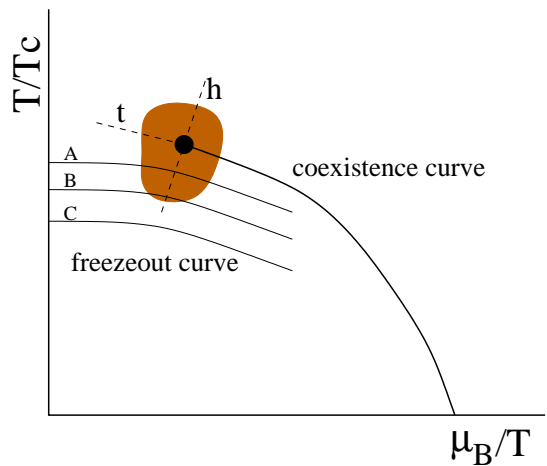}}
\end{center}
\caption{The critical point of QCD is the end point of a line of coexistence
  of two phases. One of the interesting questions for heavy-ion physics is
  how close the freezeout curve is to the critical point, \ie, which of the
  three is the true position of the freezeout curve. Our results, discussed
  below, show that $T_c=170$ corresponds to a line like A, whereas an
  increase of 5 to 10 MeV in $T_c$ could shift the results to the curve C.}
\label{fg.crit}
\end{figure}

There has been some discussion on suitable observables for the
identification of a critical point.  The first suggestion was to study the
width of the distribution of event-by-event (E/E) observations of momenta
of particles \cite{cp}. Later it was realized that a measurement which was
suggested in a different context could be suitable: that of the width of
E/E distributions of conserved quantities \cite{ebye}.  For the baryon
number, this is proportional to the baryon number susceptibility (BNS)
\cite{milc,others}, and is expected to diverge in a thermodynamically
large sample in equilibrium at the critical point. It was soon pointed
out that non-linear susceptibilities (NLS) have stronger divergence
\cite{nt4,ilgti}.  As a result, higher cumulants of the E/E distributions
are also good signals for the critical point \cite{star,caveat}.  It has
been pointed out recently that ratios of cumulants are better \cite{cpod},
since ill-determined parameters such as the fireball volume do not appear,
and they are directly comparable to predictions from lattice QCD---
\beq
  m_1 = \frac{T\chi^{(3)}(T,\mu_B)}{\chi^{(2)}(T,\mu_B)},\qquad
  m_2 = \frac{T^2\chi^{(4)}(T,\mu_B)}{\chi^{(2)}(T,\mu_B)},\qquad
  m_3 = \frac{T\chi^{(4)}(T,\mu_B)}{\chi^{(3)}(T,\mu_B)},
\label{rats}\eeq
where $\chi^{(n)}(T,\mu_B)$ is the $n$-th order NLS, obtained by taking the
$n$-th derivative of the pressure with respect to $\mu_B$. As one can see,
the ratios are not all independent, since $m_2=m_1 m_3$. These ratios are
also of interest in lattice studies of the phase diagram of QCD, since
they provide estimates of the distance to the nearest singularity of the
free energy.

At the critical point, $(T^E,\mu_B^E)$, the BNS,
$\chi^{(2)}(T^E,\mu_B^E)$, diverges. Since this critical point is in the
Ising universality class, the eigendirections of the renormalization
group scaling transformations can be classified as a ``thermal''
direction $t$ and a ``magnetic'' direction $h$, as illustrated in Figure
\ref{fg.crit}. The baryon-baryon correlation length diverges with
exponent $y_t$ as one approaches the critical point in the direction
$t$, and with exponent $y_h$ along $h$. Once these exponents are
known, the divergence of the BNS can be worked out.  In the present work
we do not need the value of the exponent $\delta$, which controls the
divergence of the baryon number susceptibility through $\chi^{(2)}(\mu_B)
\simeq |\mu_B-\mu_B^E|^\delta$.  All we need to know is that
\beq
   \chi^{(n)}\simeq |\mu_B-\mu_B^E|^{\delta+n-2}.
\label{diverge}\eeq
As a result, the ratios of successive NLS diverge as $1/|\mu_B-\mu_B^E|$
as one approaches the critical end point. Due to the CP symmetry of QCD,
a critical point at $\mu_B^E$ implies another at $-\mu_B^E$ with the same
critical exponents. Hence, the divergence of the ratios of successive
NLS can be rewritten as $1/|\mu^2-(\mu_B^E)^2|$.

In lattice calculations one uses the Maclaurin series expansion of the BNS,
\beq
   \frac{\chi^{(2)}(T,z)}{T^2} = \frac{\chi^{(2)}(T)}{T^2}
      + \frac1{2!} \chi^{(4)}(T) z^2
      + \frac1{4!} T^2 \chi^{(6)}(T) z^4
      + \frac1{6!} T^4 \chi^{(8)}(T) z^6 + {\cal O}(z^8),
\label{chi2}\eeq
organized as an expansion in the dimensionless quantity $z=\mu_B/T$ at
fixed $T$.  The Taylor (Maclaurin) coefficients, $T^{n-4}\chi^{(n)}(T)$,
which are also dimensionless, are evaluated through lattice computations
at $\mu_B=0$. We have written the expansion out to the order up to which
the coefficients are currently known \cite{nt4,nt6}. The successive
estimators of the radius of convergence of this series are
\beq
   r_{n,n+2}^2 = (n+2)(n+1)\frac{\chi^{(n)}(T)}{T^2\chi^{(n+2)}(T)}.
\label{radn}\eeq
These have been used to estimate the position of the critical point in
\cite{nt4,nt6}, where the finite volume effects in these ratios were
investigated. It was found that $r_{n,n+2}^2$ for $n\le6$ were insensitive
to the volume, provided $V\ge(4/T)^3$.

In terms of these quantities the expansion of eq.\ (\ref{chi2})
can be written as
\beq
   \frac{\chi^{(2)}(T,z)}{T^2} = \frac{\chi^{(2)}(T)}{T^2}\left[1
      + \frac{z^2}{r_{24}^2}
      + \frac{z^4}{r_{24}^2 r_{46}^2}
      + \frac{z^6}{r_{24}^2 r_{46}^2 r_{68}^2} + {\cal O}(z^8)\right],
\label{serchi2}\eeq
Clearly, the ratios of eq.\ (\ref{rats}) depend only on the three
quantities $r_{24}^2$, $r_{46}^2$ and $r_{68}^2$, and not on $\chi^{(2)}(T)$.
By taking formal derivatives of the expansion above one obtains the higher
order susceptibilities. Using these one arrives at the formal series expansions
of the measurements we are interested in---
\beqa
\nonumber
   \frac1zm_1 &=&  \frac2{r_{24}^2} 
       + z^2 2\left(\frac2{r_{46}^2}-\frac1{r_{24}^2}\right)\frac1{r_{24}^2} 
       + z^4 2\left[\frac3{r_{46}^2}\left(\frac2{r_{68}^2}-\frac1{r_{24}^2}
          \right)+\frac1{r_{24}^4}\right]\frac1{r_{24}^2}
       + {\cal O}\left(z^6\right),\\
\nonumber
  m_2 &=&  \frac2{r_{24}^2} 
       + z^2 2\left(\frac6{r_{46}^2}-\frac1{r_{24}^2}\right)\frac1{r_{24}^2} 
       + z^4 2\left[\frac1{r_{46}^2}\left(\frac{15}{r_{68}^2}-\frac7{r_{24}^2}
          \right)+\frac1{r_{24}^4}\right]\frac1{r_{24}^2}
       + {\cal O}\left(z^6\right),\\
   zm_3 &=&  1
       + z^2 \left(\frac4{r_{46}^2}\right)
       + z^4 \left(\frac3{r_{68}^2}-\frac2{r_{46}^2}\right)\frac4{r_{46}^2}
       + {\cal O}\left(z^6\right).
\label{sermeas}\eeqa
It has been demonstrated that series expansions are not a good way to
extrapolate physical quantities to finite chemical potential, and better
techniques, such as Pad\'e resummations are needed
\cite{nt6}. Since we know that $m_1$ and $m_3$ have simple poles and $m_3$
has a double pole, Pad\'e approximants of appropriate order are particularly
suited to this problem--
\beq
  m_1 = z P^L_1(z^2;a,b),\qquad
  m_3 = \frac1z P^L_1(z^2;a',b'),\qquad
  m_2 = {\widetilde P}^L_2(z^2,\tilde a,\tilde b),
\label{pade}\eeq
where the notation $P^L_M$ indicates that the numerator is a polynomial of
order $L$ and the denominator is of order $M$ and ${\widetilde P}^L_{2M}$
indicates that the denominator is the square of a polynomial of order $M$.
Since we have three terms of each series at our disposal, in each case
we can use $L=0$ and 1.  When longer series expansions become available,
one can use larger values of $L$.  The parameters $a_0\cdots a_{L-1}$,
$a'_0\cdots a'_{L-1}$, and $\tilde a_0\cdots\tilde a_{L-1}$ in the
numerator and $b_1$, $b'_1$ and $\tilde b_1$ in the denominator are
obtained by matching to the series expansions \cite{baker}. This is
done within jack-knife blocks.  We check that the relation $m_2=m_1m_3$
is satisfied within errors.

\begin{table}[tbh]
\begin{center}
\begin{tabular}{|c|c|c|c|c|}
\hline
$\snn$ & $z$ & $T/T_c$ & $\beta(N_t=6)$ & $\beta(N_t=4)$ \\
\hline
  5.0 & 4.7   (2) & 0.70 (3) &  & 5.20 \\
  7.7 & 3.02  (1) & 0.82 (2) &  & 5.22 \\
 11.5 & 2.08  (7) & 0.89 (1) & 5.39 & 5.24?\\
 18.0 & 1.39  (5) & 0.94 (1) & 5.41 & 5.275\\
 19.6 & 1.29  (4) & 0.94 (1) & 5.41 & 5.275\\
 27.0 & 0.96  (2) & 0.96 (1) & 5.415& 5.275 \\
 39.0 & 0.68  (2) & 0.97 (1) & 5.415& 5.275\\
 62.4 & 0.44  (2) & 0.97 (1) & 5.415& 5.275\\
200.0 & 0.142 (5) & 0.98 (1) & 5.415& 5.275\\
850.0 & 0.034 (1) & 0.98 (1) & 5.415& 5.275\\
2500.0 & 0.0115 (4) & 0.98 (1) & 5.415& 5.275\\
5000.0 & 0.0058 (2) & 0.98 (1) & 5.415& 5.275\\
\hline
\end{tabular}
\end{center}
\caption{The values of $z=\mu_B/T$ and $T/T_c$ at freezeout for various
$\snn$ are shown for $T_c=170$ MeV. The numbers in brackets are errors in
the least significant digit and come from the statistical error quoted in
\cite{freezeout}. There are also statistical errors on the scale setting
on the lattice \cite{nt4,nt6}. The combination of these errors may result
in the same coupling, $\beta$, corresponding to several different central
values of $T/T_c$. Missing entries for $\beta$ mean that the simulations
in \cite{nt4,nt6} do not correspond to these conditions).}
\label{tb.freeze}
\end{table}

In order to compare lattice predictions with experiment one needs the
values of $T$ and $\mu_B$ along the freezeout curve as a function of
the center of mass energy of the colliding nuclei, $\snn$.
These were extracted from data in \cite{freezeout}. For a comparison
with lattice computations it is necessary to convert to the dimensionless
variables $z=\mu_B/T$ and $T/T_c$. We have used a crossover temperature
$T_c=170$ MeV consistent with the current best estimates \cite{rbrc,wubu}.
The correspondences between lattice parameters and the freezeout points
are given in Table \ref{tb.freeze},  where we have tabulated parameters
corresponding to the RHIC low-energy run as well as the LHC heavy-ion
runs at both full energy and the expected energy at the end of the
current year.

Once all non-thermal sources of fluctuations are eliminated from
data, the ratios of eq.\ (\ref{rats}) are directly comparable to
lattice QCD predictions. Lattice predictions have long been used to
interpret experimental data from heavy-ion collisions; examples being
the qualitative agreement of the equation of state of hot QCD matter
with experimentally extracted values and the agreement of strangeness
production parameters extracted from lattice measurements and fits to
hadron yields. These have been indirect agreements, since a layer
of analysis stands between the data and the prediction. The main
results of this paper allow direct comparison of lattice predictions and
data for these ratios along the freezeout curve. Agreement of data with
these values are direct tests of lattice QCD, and could potentially be the
first quantitative tests of lattice QCD in a thermal environment. However
it is interesting to point out other aspects of this argument.

If the critical point is far from the freezeout curve over a certain
range of energy, then $m_1$ decreases with increasing $\snn$
(since $z$ decreases) and $m_3$ increases.  Using these two measurements
and comparing with lattice predictions, it is possible to estimate
the freezeout conditions: $T/T_c$ and $\mu_B/T$. This method is
independent of the usual one in which hadron yields are interpreted
through a resonance gas picture \cite{freezeout}. Comparison of the
two methods then allows us to estimate $T_c$ by inverting the argument
of the previous paragraph.  Mutual agreement of the values of $T_c$
so derived at different $\snn$ would constitute the first firm
experimental proof of thermalization. If this proof holds then one also
obtains the simplest and most direct measurement of $T_c$ found till now.
Since such a thermometric measurement can be made reliably with data at
large $\snn$, where $\mu_B$ is small, it would remain a valid
measurement whether or not a critical point is found in the low energy
scan at RHIC.

Our numerical results are based on the measurements of the Taylor
coefficients in the expansion of eq.\ (\ref{chi2}) which were reported
in \cite{nt4,nt6}. Since the data were taken at two different series
of lattice spacings, with $N_t=4$ and 6, one can examine the approach
to the continuum limit.  In both sets of simulations the pion mass was
close to physical, being about 230 MeV. The spatial volume varied from
$(2/T)^3$ to $(4/T)^3$ at all lattice spacings, and up to $(6/T)^3$
on the coarser lattice.  The present analysis was performed using a
jack-knife estimator of the mean and its error with 5--10 jack-knife
blocks. We used only statistically independent measurements, of which
we have more than 50 at each coupling and volume.

A test of the method is to use it at the previously determined value of
$T^E/T_c$ and check whether the extrapolations give radii of convergence
which are compatible with that previous estimate. The
Taylor expansion of $\chi^{(2)}$ around $z_0=\mu_B^0/T$ is
\beq
   \frac{\chi^{(2)}(T,z)}{T^2} = \frac{\chi^{(2)}(T,z_0)}{T^2}
   + \frac{\chi^{(3)}(T,z_0)}{T} (z-z_0)
   + \frac{\chi^{(4)}(T,z_0)}{2!} (z-z_0)^2
   + {\cal O}\left((z-z_0)^3\right).
\label{taylor}\eeq
This implies that $1/m_1$ and $2/m_3$ are estimates of the radii
of convergence of the series. When they are computed at $T^E$
then $z_0+1/m_1$ and $z_0+2/m_3$ have to be compatible with previous
estimates of $\mu^E/T^E$, obtained from the Maclaurin expansion of eq.\
(\ref{serchi2}).  We have verified this for both $N_t=4$ and 6.

This argument also allows us to understand systematics due to
lattice volume effects and finite lattice spacing, and, therefore,
the possible effects of improved actions. Away from the vicinity of the
critical point finite volume effects are expected to be small as long
as $VT^3\ge4^3$. Finite volume effects are expected to be larger near
the critical point: indeed this is one way to check whether lattice
extrapolations show genuine critical effects. Lattice spacing effects
are also easy to understand. Since the Maclaurin series for $zm_3$
starts with unity, lattice spacing correction are expected to be
negligible and small. Indeed, this is what is observed. On the other
hand, since the Maclaurin series for $m_1/z$ and $m_2$ start with the
terms $\chi^{(4)}(T)/\chi^{(2)}(T)$, one expects that a large part of
these lattice effects can be subsumed into the lattice spacing effects
on these quantities. We can correct for the lattice spacing changes
between $N_t=4$ and 6 using data from \cite{nt4,nt6}.  However, there
is not yet enough data to extrapolate to the continuum.  Note however,
that $m_1/z$ and $m_2$ have a common normalization. Hence we normalize
to $m_2=1$ at $\snn=5$ TeV.

\begin{figure}[t!]
\begin{center}
\scalebox{0.6}{\includegraphics{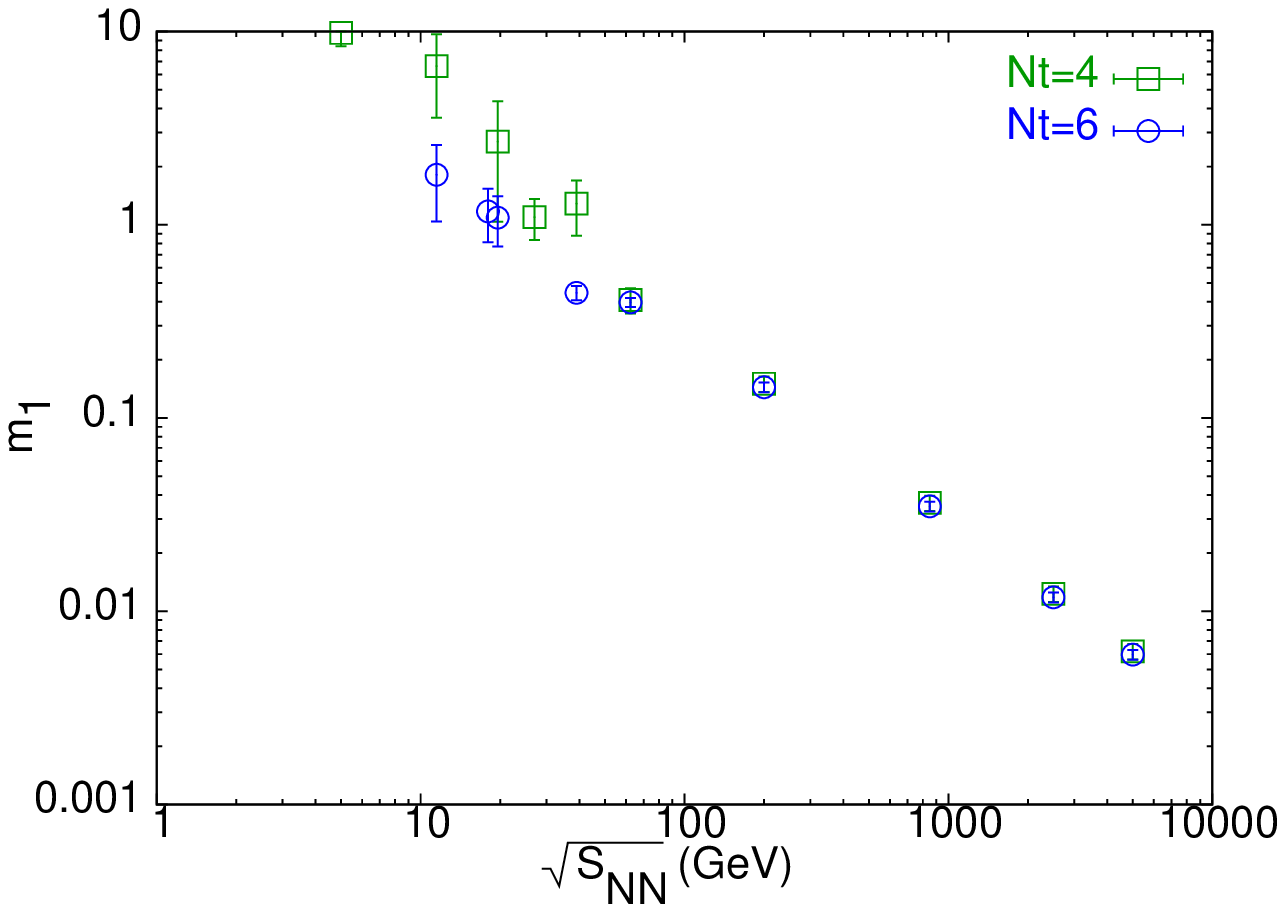}}
\scalebox{0.6}{\includegraphics{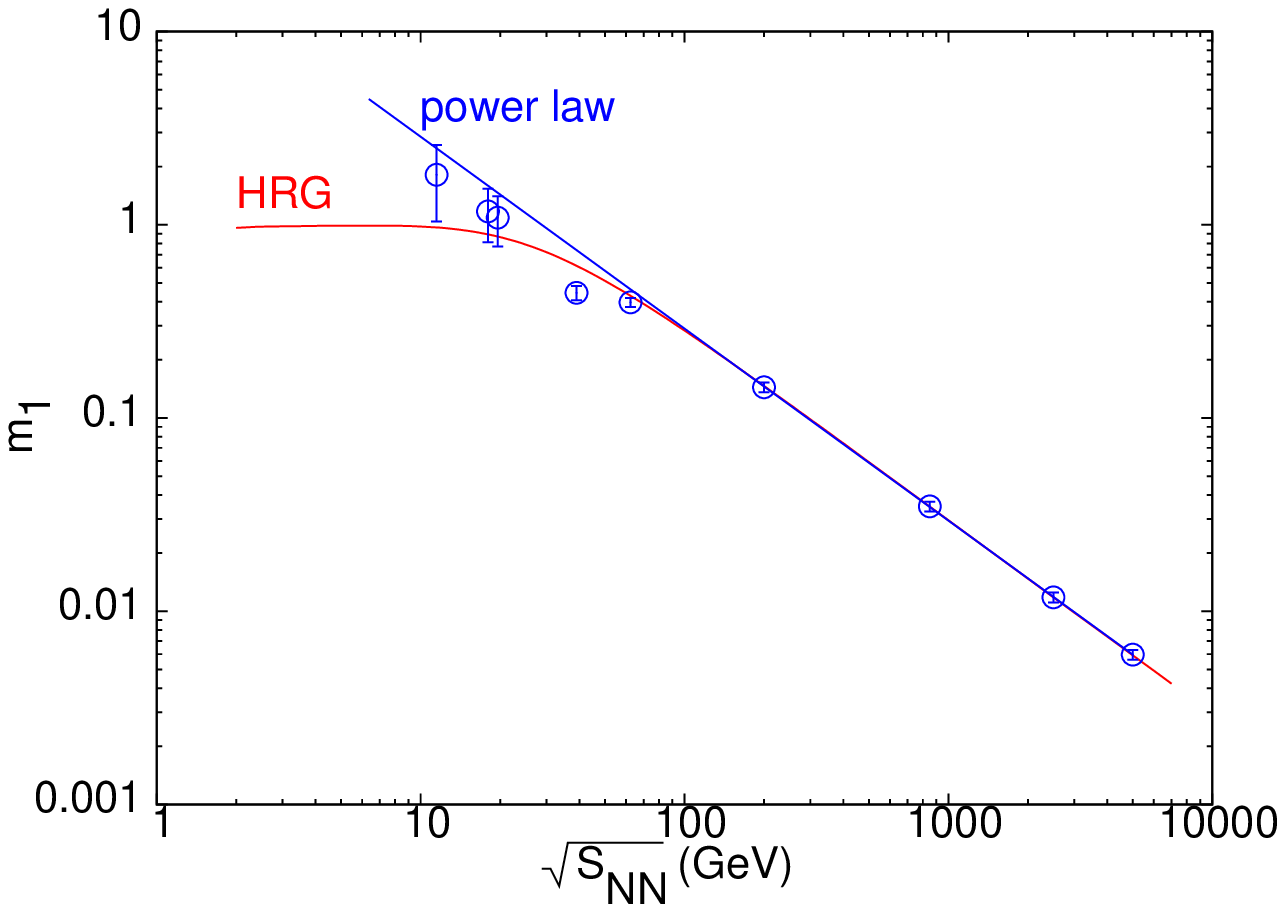}}
\scalebox{0.6}{\includegraphics{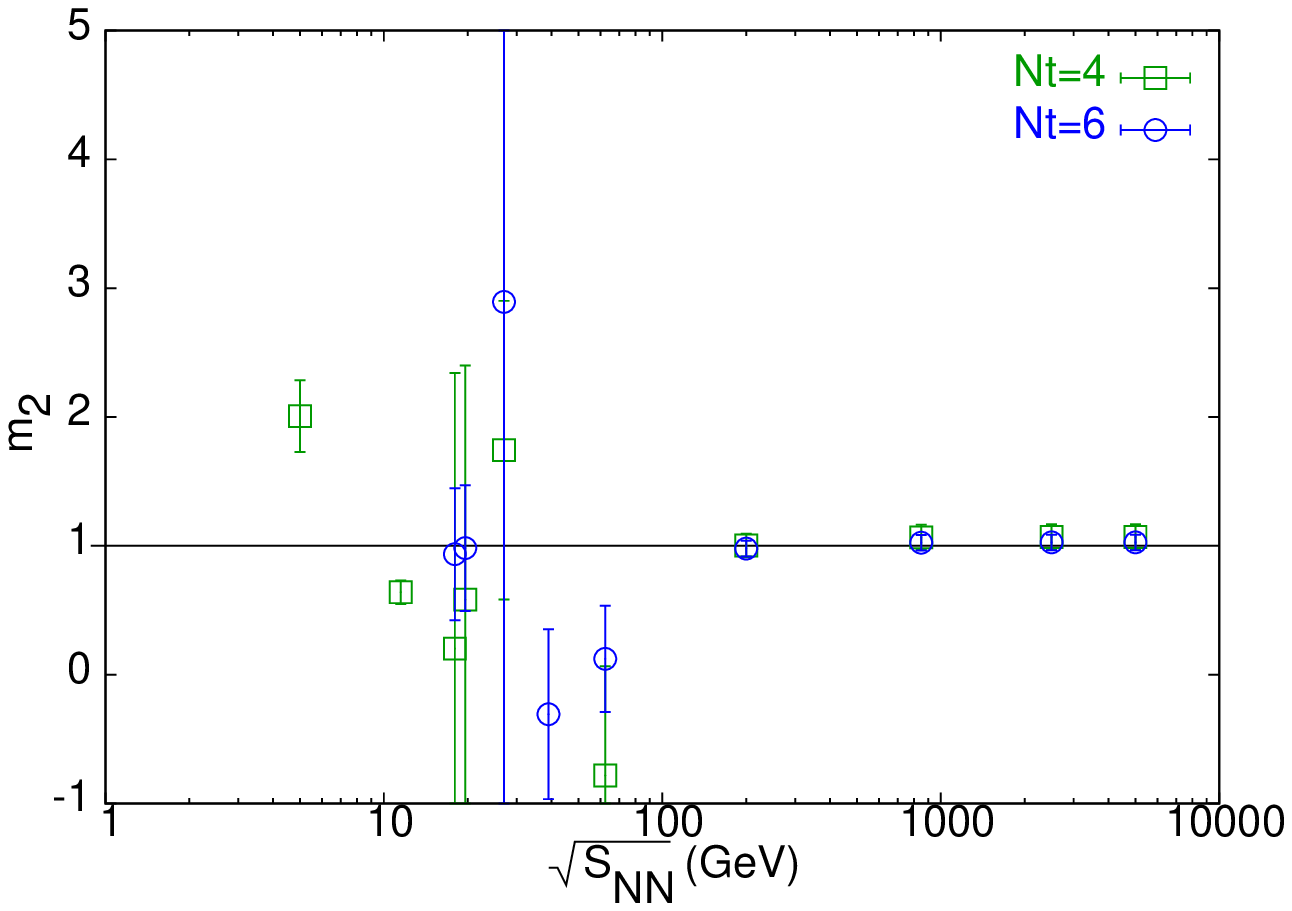}}
\scalebox{0.6}{\includegraphics{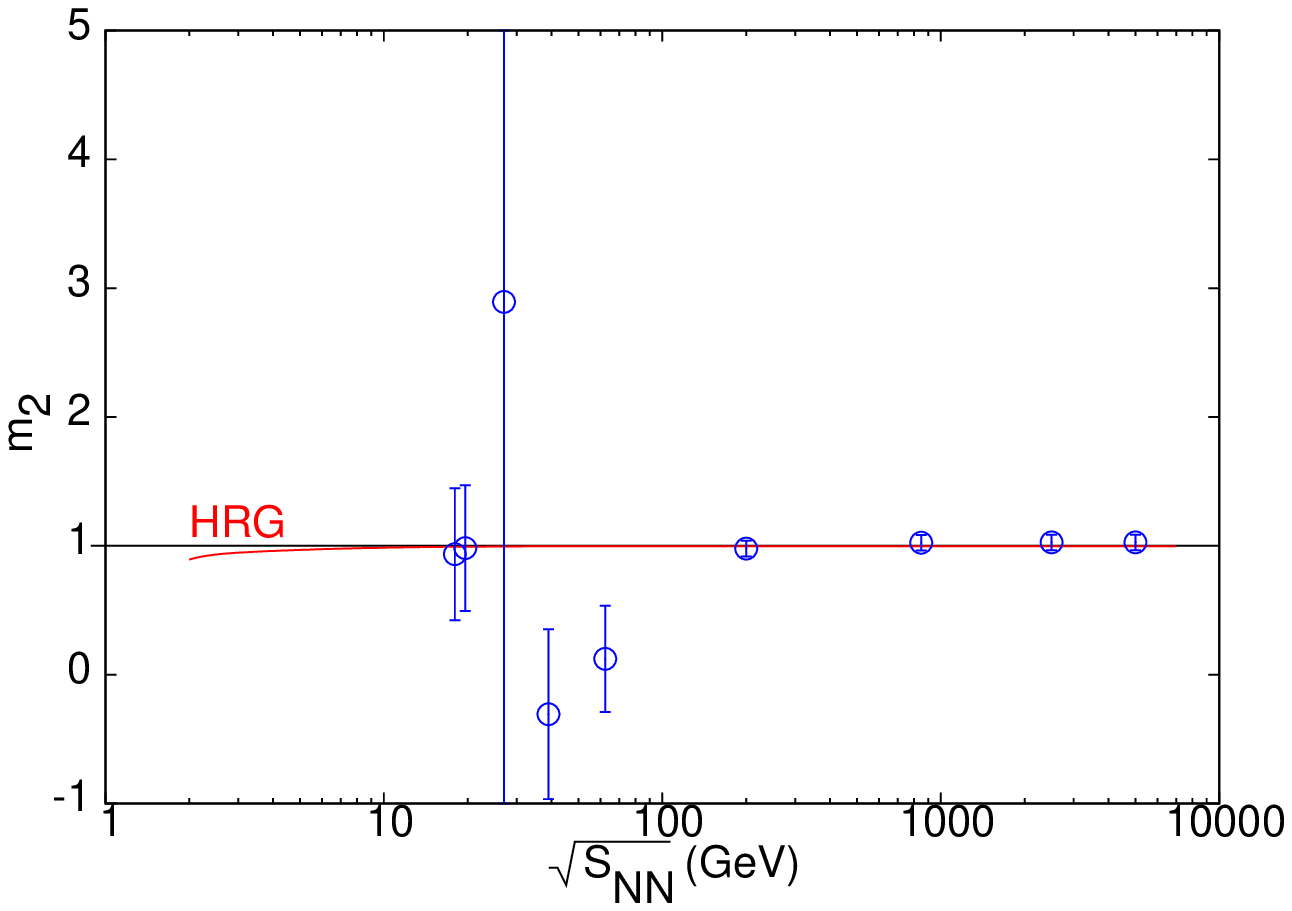}}
\scalebox{0.6}{\includegraphics{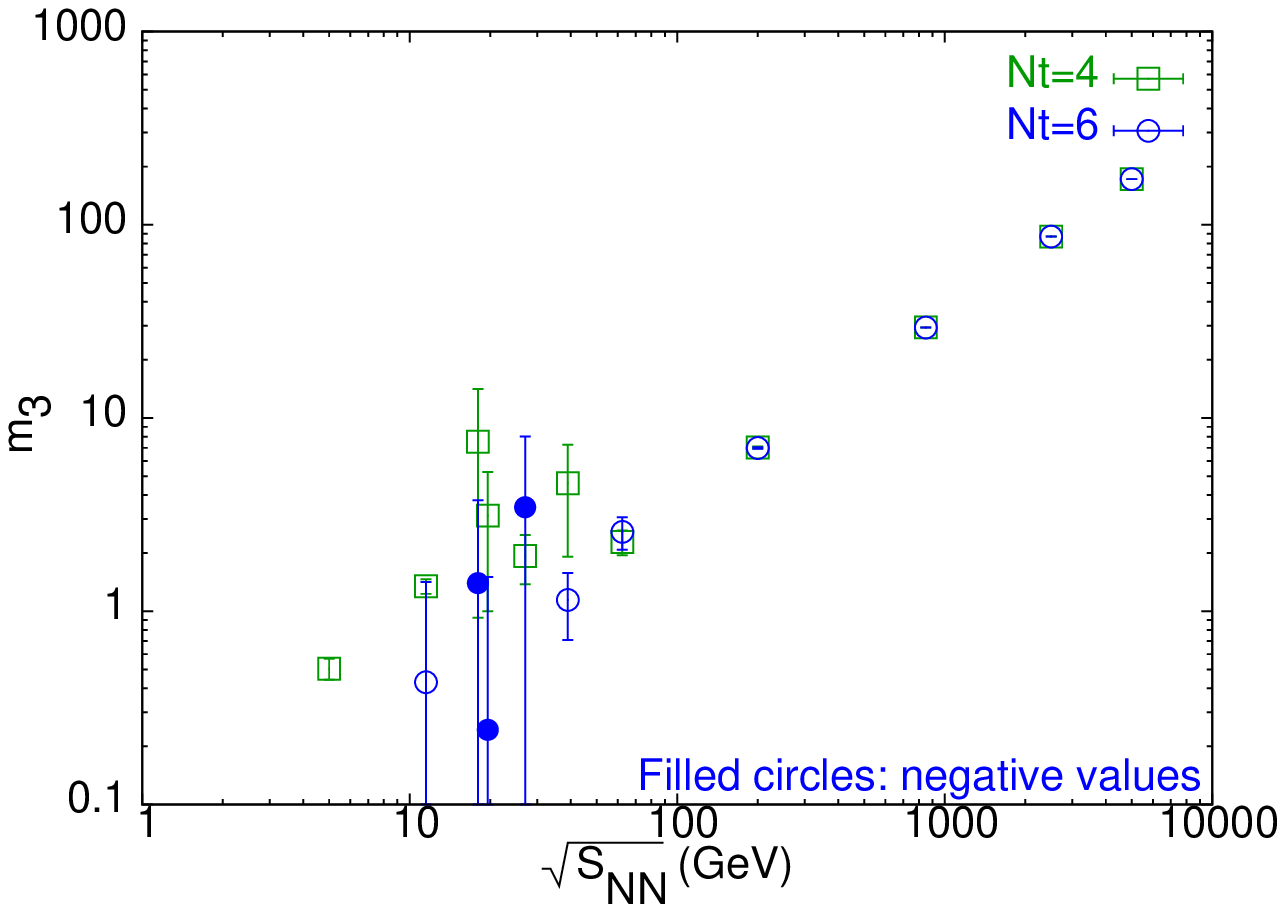}}
\scalebox{0.6}{\includegraphics{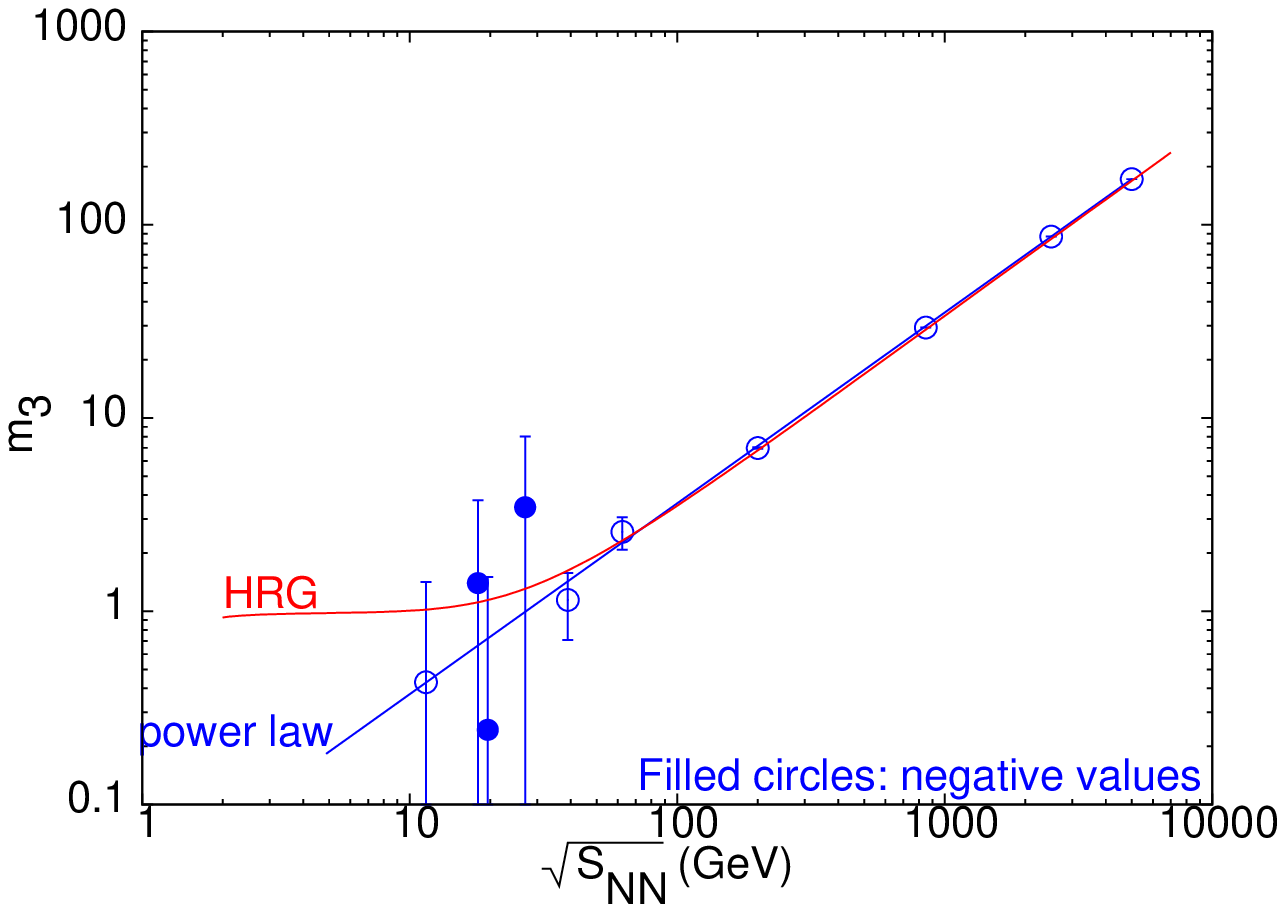}}
\end{center}
\caption{The ratios $m_1$, $m_2$ and $m_3$ along the freeze out curve as
 a function of $\snn$. The panels on the left compare the lattice results
 for $N_t=4$ and 6. Those on the right show only $N_t=6$ along with the
 hadron resonance gas results \cite{sandeep}, marked HRG, and a simple
 power law extrapolation of high-energy data, in order to make clearer
 the non-monotonicity near the critical region.}
\label{fg.limb}
\end{figure}

With these normalizations, we find that at freezeout in the highest RHIC
energy of $\snn=200$ GeV
\beq
   m_1 = 0.140 \pm 0.008, \quad
   m_2 = 0.95 \pm 0.06, \quad
   m_3 = 6.99 \pm 0.06.
\label{pred}\eeq
The errors shown are purely statistical. The remaining systematic
error comes from the imprecision in our current knowledge of $T_c$. We
can estimate the magnitude of this systematic error by changing $T_c$
by 5 MeV. For example, using $T_c=175$ we find, using the corresponding
normalization, that 
\beq
   m_1 = 0.14 \pm 0.01, \quad
   m_2 = 1.04 \pm 0.07, \quad
   m_3 = 7.3  \pm 0.1.
\label{predold}\eeq
Hence this source of uncertainty changes results by less than 10\% at
the highest RHIC energy. Closer to the critical region the changes can
be much larger. For instance, whether $m_2$ and $m_3$ are negative at
some $\snn$, as seen in Figure \ref{fg.limb}, depends sensitively on
the choice of $T_c$--- not being visible for the higher value of $T_c$.

The dependence of $m_1$, $m_2$ and $m_3$ on the energy, $\snn$,
along the freezeout curve is shown in Figure \ref{fg.limb}. Note that the
lattice spacing dependence is under good control for $\snn>50$
GeV. In this region of energies $z=\mu_B/T$ is small, as a result of which
the Pad\'e approximants are almost constant and one has $m_1\simeq z$,
$m_2\simeq1/z$ and $m_3\simeq z^0$.  Since the freezeout temperature
here is smaller than $T_c$, one should compare our results with the
predictions of resonance gas models. These give $m_2=1$ \cite{sandeep,reson},
which is close to our results. In our normalization an ideal quark gas
yields $m_2=2/(3\pi^2)$, which is very different.

In the region of energy just below $\snn=50$ GeV the lattice
spacing dependence of the results is not under good control yet. This
is related to the fact that the critical point observed on the lattice
\cite{nt4,nt6} lies in this region and shifts with the lattice spacings
used. Due to this shift in the critical point with lattice spacing,
the two lattice spacings give different predictions in this region. As
mentioned above, the choice of $T_c$ is also important. In
order to show the expected behaviour near the critical point, we have
also shown the results for a single lattice spacing--- the smallest
currently available. Note that the behaviour of $m_1$ is fairly smooth,
although a departure from the behaviour $m_1\simeq z$ is clear. Both
$m_2$ and $m_3$ show signs of non-monotonic behaviour with the Kurtosis
becoming negative. Such leptokurtic behaviour is impossible in the hadron
resonance gas. It develops in QCD only when the lattice spacing is small
enough.

At very small $\snn$ one again leaves the vicinity of the critical
point, and the lattice spacing effects are again under reasonable control.
This happens although $z=\mu_B/T$ is very much larger than unity,
and the lattice measurements, made at $z=0$, have to be extrapolated
very far. The reliability of the lattice prediction at large $z$ is due
to the stability of the Pad\'e resummations. Note that in this region
there is a return to the expected behaviour $m_1\simeq z$,
$m_2\simeq1/z$ and $m_3\simeq z^0$.

Since thermodynamic stability requires $\chi^{(2)}$ to be positive,
the signs of $m_1$ and $m_2$ immediately tell us the signs of  both
$\chi^{(3)}$ and $\chi^{(4)}$.  Our observations then show that $\chi^{(3)}$
is positive along the freezeout curve, in agreement with the qualitative
argument of \cite{caveat}. This implies that those parts of the freezeout
curve which current lattice computations can access lie entirely below
the line of first-order transitions.  It is a moot question whether the
freezeout line crosses the first-order line (\ie, the coexistence line
of Figure \ref{fg.crit}) at lower energies or whether future lattice
computations at smaller lattice spacings will show different behaviour.

The most pleasant aspect of the current results is that the critical
point is close enough to the freezeout curve for its effect to be visible
as non-monotonic behaviour of the ratios. This is clearest for $m_1$
because of the relatively smaller error bars. However, the effect is
also present in $m_2$ and $m_3$.  There is currently some uncertainty
in the position and size of the peak due to uncertainties in lattice
determinations of $T_c$. However, all the available lattice data show
that the critical point lies within reach of the proposed energy scan
at the RHIC.  We emphasize that the existence of non-monotonicity is a
general expectation if there is a nearby critical point. The baseline
over which the behaviour stands is a precise result from lattice QCD
which is new.

In this letter we have examined lattice QCD predictions for the ratios
in eq.\ (\ref{rats}). On the lattice they are related to the radius of
convergence of the series in eq.\ (\ref{taylor}), and they are also
measurable in experiments \cite{cpod}. Using Pad\'e approximants to
resum the series expansion of these ratios (which reproduce previous
results on the location of the critical point) we predict the ratios
along the freezeout curve (see Figure \ref{fg.limb}).  Current lattice
results already show the following important qualitative and quantitative
trends---
\begin{enumerate}
\item Generally $m_1$ decreases with increasing $\snn$ and $m_3$
  increases, whereas $m_2$ has a flatter dependence on energy. This general
  trend is interrupted in the neighbourhood of a critical point where there
  are rapid and non-monotonic changes in all three observables.
\item At the highest energies (RHIC and LHC) sufficiently accurate measurements
  of $m_1$ and $m_3$ may be used to deduce the value of the crossover
  temperature, $T_c$.
\item The skew and kurtosis are both expected to be positive in the range of
  energies away from the critical point, with a possible change in the sign
  of the kurtosis in the critical region.
\end{enumerate}

These computations were performed on the Cray X1 of the ILGTI at TIFR.
RVG would like to acknowledge the hospitality of the Department of Physics
at the University of Bielefeld. SG would like to thank Bedangadas Mohanty
and Nu Xu for discussions. We would like to thank Sandeep Chatterjee for
tables of $m_1$, $m_2$ and $m_3$ along the freezeout curve in the hadron
resonance gas model.

\end{document}